\pdfoutput=1

\documentclass[twocolumn,showpacs,aps,prl,superscriptaddress]{revtex4-1}

\usepackage{graphicx}
\usepackage{upgreek}

\usepackage{bm}
\usepackage{amssymb}
\usepackage{amsmath}
\usepackage{subfigure}

\usepackage{color}

\renewcommand{\text}[1]{\ensuremath{\textrm{#1}}}

\DeclareMathOperator\erf{erf}
\begin{document}

\title{Near-field interferometry of a free-falling nanoparticle from a point-like source}

\author{James Bateman}
\affiliation{School of Physics and Astronomy, University of Southampton, Highfield, SO17 1BJ, UK}
\author{Stefan Nimmrichter}
\affiliation{Fakult\"at f\"ur Physik, Universit\"at Duisburg-Essen, 47048 Duisburg, Germany}
\author{Klaus Hornberger}
\affiliation{Fakult\"at f\"ur Physik, Universit\"at Duisburg-Essen, 47048 Duisburg, Germany}
\author{Hendrik Ulbricht}\email{h.ulbricht@soton.ac.uk} 
\affiliation{School of Physics and Astronomy, University of Southampton, Highfield, SO17 1BJ, UK}

\date{\today}

\newcommand{\tr}[1]{\text{tr}\left( #1 \right)}

\newcommand{\eps}{\varepsilon}
\newcommand{\vareps}{\varepsilon}
\newcommand{\id}{\text{id}}
\newcommand{\la}{\langle}
\newcommand{\ra}{\rangle}
\newcommand{\da}{{\dagger}}

\newcommand{\V}{ {\cal V} }

\newcommand{\re}{\text{Re}}
\newcommand{\im}{\text{Im}}
\newcommand{\sgn}{\text{\rm sgn}}
\newcommand{\sinc}{\text{\rm sinc}}
\newcommand{\Si}{\text{\rm Si}}
\renewcommand{\erf}{\text{\rm erf}}
\newcommand{\const}{\text{const}}

\begin{abstract}
Matter-wave interferometry performed with massive objects elucidates their wave nature and thus tests the quantum superposition principle at large scales.
Whereas standard quantum theory places no limit on particle size, alternative, yet untested theories---conceived to explain the apparent quantum to classical transition---forbid macroscopic superpositions.
Here we propose an interferometer with a levitated, optically cooled, and then free-falling silicon nanoparticle in the mass range of one million atomic mass units, delocalized over more than 150 nm.
The scheme employs the near-field Talbot effect with a single standing-wave laser pulse as a phase grating.
Our analysis, which accounts for all relevant sources of decoherence, indicates that this is a viable route towards macroscopic high-mass superpositions using available technology.

\end{abstract}


\maketitle

\section{Introduction}
Matter-wave interference with particles of increasing size and mass is a natural and viable method for testing the validity of the quantum superposition principle at unprecedented macroscopic scales \cite{Clauser1997,Leggett2002,hornberger2012colloquium,bassi2013models}. Macroscopic path separations are nowadays routinely achieved in atom interferometry \cite{Mueller2008,Muentinga2013,Dickerson2013}, and technological advances in the control of opto-mechanical systems \cite{aspelmeyer2013cavity} promise that 
much more massive objects may be delocalized \cite{marshall2003towards,chang2010cavity,romero-isart2010toward,scala2013matter,yin2013large}, 
albeit with spatial separations smaller than a single atom. 

Recent proposals put forward nanoparticle interferometry \cite{romero-isart2011large,nimmrichter2011concept} in the mass range of $10^6$ to $10^9\,$amu to surpass the mass records currently held by molecule diffraction experiments \cite{hornberger2012colloquium,eibenberger2013matter}, while maintaining spatial separations large enough to be resolved by optical means. 
A first demonstration with molecular clusters \cite{haslinger2013universal} is still far away from the mentioned high mass regime due to hard experimental challenges, mainly concerning the source and detection. The realization of a proposed double-slit scheme with silica nanospheres \cite{romero-isart2011large} requires motional ground state cooling, which is an equally challenging task. 

Quite recently, optical feedback cooling has been demonstrated for $100\,$nm-sized particles 
\cite{li2011millikelvin, gieseler2012subkelvin},
based on pioneering work that demonstrated the trapping of polystyrene and glass microspheres~\cite{ashkin1986observation}, trapping of viruses and bacteria~\cite{ashkin1987virusesbacteria} and even of complete cells~\cite{ashkin1987optical} in solutions and high-vacuum. Cavity cooling of particles of similar size was proposed~\cite{barker2010cavity} and recently achieved \cite{kiesel2013cavity,asenbaum2013cavity} in one dimension, with temperatures in the milli-Kelvin range. Although this is still far above the ground state of a typical $100\,$kHz trap, we will argue that high-mass interference can be realized experimentally with motional temperatures already achieved by optical cooling.

In this Letter we present a near-field interference scheme for $10^6$\,amu particles. It is based on the single-source Talbot effect \cite{brezger2003concepts} due to a single optical phase grating, as opposed to the three-grating scenario in Talbot--Lau interference experiments \cite{hornberger2012colloquium}. Optically trapped silicon nanospheres, feedback-stabilized to a thermal state of about $20\,$mK, provide a sufficiently coherent source. Individual particles are dropped and diffracted
by a standing UV laser wave, such that interference of neighboring diffraction orders produces a resonant near-field fringe pattern. In order to record the interferogram, the nanospheres are deposited on a glass slide and their arrival positions are recorded via optical microscopy. We argue that the choice of silicon, due to its specific material characteristics,  will yield reliable high mass interference, unaffected by environmental decoherence, in a setup that can be realized with present-day technology.

\begin{figure}[b]
\centerline{\includegraphics[width=0.75\columnwidth]{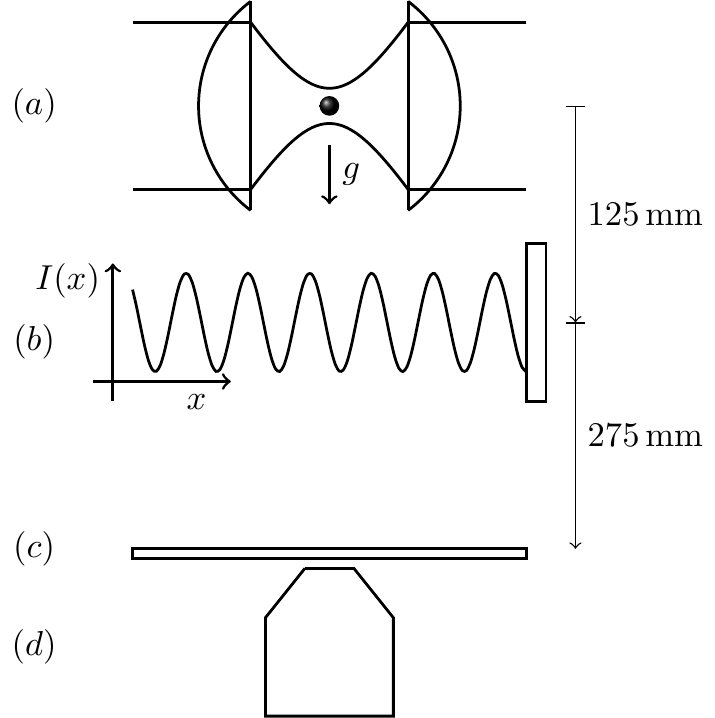}}
\caption{(a) Laser light at 1550\,nm
creates a harmonic trap for single silicon nanospheres. After feedback cooling to 20\,mK, the particle is released and falls for 125\,mm,
where it passes a phase grating (b) provided by a retro-reflected nanosecond pulse at 355\,nm. 275\,mm further down the particle is adsorbed on a glass slide (c), where the arrival position is recorded with 100\,nm accuracy via optical microscopy
(d).}
\label{fig:schematic}
\end{figure}

\section{Results}
\subsection{Proposed Experiment}\label{sec:proposed-experiment}
The proposed scheme is sketched in Fig.~\ref{fig:schematic}. In the first stage of the experiment, a silicon particle is captured in an optical dipole trap by a lens system of numerical aperture $0.8$ focusing a $1550\,$nm laser to a waist of $860\,$nm~\cite{richards1959electromagnetic}; the interaction of nanoparticles with light is described further in Supplementary Note 1.
The trapping light is collected and used to determine the position of the particle~\cite{gieseler2012subkelvin}, which is feedback cooled over many trapping cycles to about $T=20\,$mK of mean translational energy along the horizontal $x$-axis, implying a momentum uncertainty $\sigma_p=\sqrt{m k_{ B}T}$ of about $\sigma_p/m=1.2$\,cm/s. A laser power of $55\,$mW results in a trap frequency of $\nu_{\rm M}=200\;$kHz and a position  uncertainty $\sigma_x$ below 10\,nm; see Supplementary Note 2.
The trap thus serves as a nearly point-like matter-wave source for diffraction.

\begin{figure*}[tb]
\includegraphics{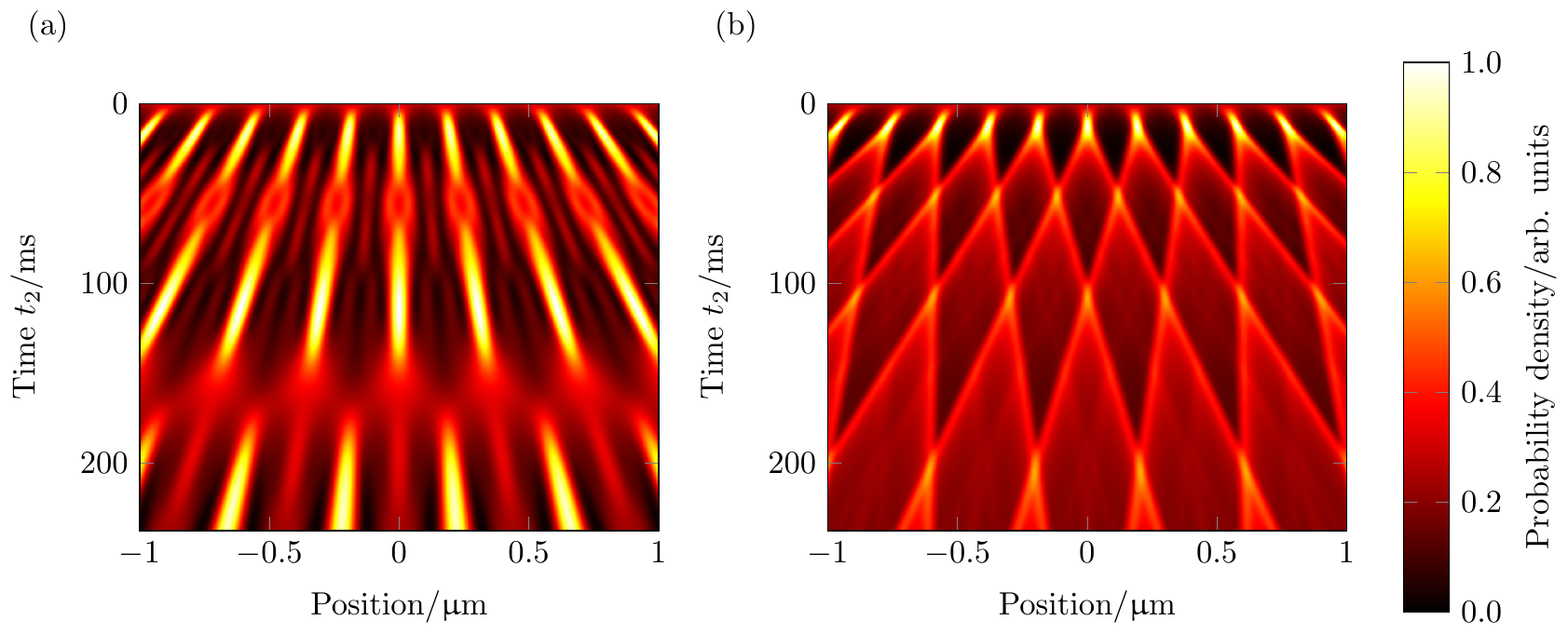}
\caption{(a) Quantum and (b) classical probability densities at different times $t_2$ after the grating pulse.
We assume the width of the trap state to be $\sigma_x=10\;$nm and a time of flight to the grating of $t_1=160$\,ms; the grating has the period  $d=177\;$nm with the maximal phase-modulation set to $\phi_0=\pi$. 
The interference pattern is detected after $t_2=126$\,ms in the setup of Fig.~1. 
Quantum mechanics then predicts high contrast fringes which cannot be explained classically.}
\label{fig:NormalCarpets}
\end{figure*}

After feedback cooling, the particle is released from the trap and falls for $t_1=160\,$ms before it is illuminated by a frequency-trippled Nd:YAG laser pulse at $355\;$nm with a pulse length of $10\;$ns and an energy $E_{\rm G} \leq 500\,\upmu\text{J}$.
The pulse is retro-reflected by a mirror to form a standing-wave phase grating with period $d=\lambda_{\rm G} /2$, which diffracts the particle by modulating the matter-wave phase through the dipole interaction.
The Talbot time, which sets the scale for near-field interference \cite{hornberger2012colloquium}, is thus given by $t_{\rm T}=md^2/h\approx 80$\,ms.
The laser beam must be expanded such that the waist is larger than the uncertainty in position $\sigma_pt_1/m\approx 2\,$mm accrued during free-flight.
Moreover, the orientation of the grating must be angularly stable to less than micro-radians to avoid blurring of the interferogram due to acceleration of the particle under gravity, and positionally stable to within $30\,$nm relative to the initial particle position; see Supplementary Note 3.

After the grating, the particle undergoes free-fall for $t_2=126\,$ms forming an interference pattern when it arrives on the glass slide. The arrival position can be detected by absorption imaging with visible light. Fitting to the known point-spread function of the imaging system permits 100~nm positional accuracy~\cite{hell2007far}; see Supplementary Note 4.
The density pattern depicted in Fig.~\ref{fig:NormalCarpets}(a) is predicted to appear after many runs of the experiment. In the following, we discuss the theoretical description of the interference effect and the experimental constraints.

\subsection{Theoretical model}
Our starting point for evaluating the interference effect is the trapped thermal state of motion, a Gaussian mixture with standard deviations $\sigma_x = \sqrt{k_B T / 4\pi^2 m\nu_{\rm M}^2}$ and $\sigma_p = \sqrt{m k_B T}$. 
The particle will be illuminated by a uniform standing-wave pulse oriented along the horizontal $x$-axis (see Fig.~\ref{fig:schematic}), so that the $y$- and $z$-motion can be ignored. 

The near-field diffraction effect including all relevant decoherence mechanisms 
is best captured in a quantum phase-space description 
\cite{hornberger2004theory}. For the present purposes  it is most useful to work with the characteristic function representation $\chi \left( s,q\right)$, i.e.~the Fourier transform of the Wigner function  \cite{schleich2001quantum} of a given quantum state $\rho$. Here, we summarize the detailed derivation given in the Supplementary Methods.

The initial Gaussian state,
\begin{equation}
 \chi_0 \left(s,q\right) = \exp\left(-\frac{\sigma_{x}^{2}q^{2}+\sigma_{p}^{2}s^{2}}{2\hbar^{2}}\right), \label{eq:chi_trap}
\end{equation}
first evolves freely for a time $t_1$, $\chi_{1} \left(s,q\right) = \chi_0 \left( s-qt_1 /m,q \right)$, before it is illuminated by the optical grating pulse of period $d$. Given an almost point-like initial spread $\sigma_x/d \ll 1$, the matter waves must evolve for at least the Talbot time $t_{\rm T} $,
to ensure that they are delocalized over adjacent grating nodes in order to be able to interfere. The initial momentum, on the other hand, is spread over many grating momenta, $\sigma_p d / h \gg 1$, so that the time-evolved state extends over many grating periods. That is, if particles are only detected in a finite detection window around the center of the distribution in the end, we can neglect the Gaussian density profile by writing
\begin{equation}
 \chi_{1} \left(s,q\right) \approx \frac{\sqrt{2\pi}\hbar}{\sigma_p}\exp\left(-\frac{\sigma_{x}^{2}q^{2}}{2\hbar^{2}}\right) \delta \left( s - \frac{q t_1}{m} \right). \label{eq:chi_t1}
\end{equation}
The particle interacts with the standing-wave pulse through its optical polarizability $\alpha = 4\pi\eps_0 R^3 \left(n_{\rm Si}^2 - 1\right)/\left(n_{\rm Si}^2 + 2\right)$, determined by the particle radius $R$ and its complex refractive index $n_{\rm Si}$ at the grating wavelength $\lambda_{\rm G}=2d$. In the limit of short pulse durations $\tau$, this imprints the phase $\phi(x)=\phi_0 \cos^2 \left(\pi x/d\right)$ on the matter-wave state \cite{nimmrichter2011concept}, where $\phi_0 = 2\re\left( \alpha\right) E_{\rm G} / \hbar c\eps_0 a_{\rm G}$ depends on the energy $E_{\rm G}$ and spot area $a_{\rm G}$ of the pulse. The characteristic function transforms as $\chi_{1} \left(s,q\right) \to \sum_n B_n \left( s/d\right) \chi_{1} \left( s,q+nh/d \right)$, where the $B_n$ are Talbot coefficients, given in terms of Bessel functions \cite{hornberger2009theory}, 
\begin{equation}
 B_n \left( \xi \right) = J_n \left( \phi_0 \sin \pi\xi \right). \label{eq:Bn}
\end{equation} 
Incoherent effects due to absorption or scattering of laser photons are negligible for the nanoparticles considered here (Supplementary Methods); 
nevertheless, our numerical simulations include both effects.

The final density distribution $w(x)=\la x|\rho|x\ra$, i.e.~the probability to find the particle at position $x$ after another free time evolution by $t_2$, then takes the form
\begin{eqnarray}
 w(x) &=& \frac{m}{\sqrt{2\pi}\sigma_p \left( t_1+t_2 \right)} \sum_n B_n \left[ \frac{nt_1 t_2}{t_{\rm T} \left( t_1+t_2 \right)} \right] \nonumber \\
 &&\times \exp \left[ \frac{2\pi i n x}{D} - \frac{2\pi^2 n^2 \sigma_x^2 t_2^2}{d^2 \left( t_1+t_2 \right)^2} \right]. \label{eq:pattern}
\end{eqnarray}
It describes a periodic fringe pattern oscillating at the geometrically magnified grating period $D = d \left( t_1 + t_2 \right)/t_1$ \cite{brezger2003concepts}. The fringe amplitudes, given by the Talbot coefficients (\ref{eq:Bn}),
are diminished the larger the spread $\sigma_x$ of the initial state (\ref{eq:chi_trap}). 

An exemplary density pattern (\ref{eq:pattern}) is plotted in Fig.~\ref{fig:NormalCarpets}(a) for varying time $t_2$. The simulation was performed for $10^6\,$amu silicon particles, assuming realistic experimental parameters 
and including the influence of environmental decoherence.
It shows pronounced interference fringes with visibilities of up to 75\%. 

The pattern in Fig.~\ref{fig:NormalCarpets}(b) is the result of a classical simulation assuming that the particles are moving on ballistic trajectories. A lensing effect due to the strong dipole forces exerted by the standing-wave field is here responsible for the density modulation. This classical result is obtained simply by replacing $\sin\pi\xi$ by $\pi\xi$ in the expression (\ref{eq:Bn}) for the grating coefficients \cite{hornberger2009theory}.

\begin{figure}
  \centering
\includegraphics{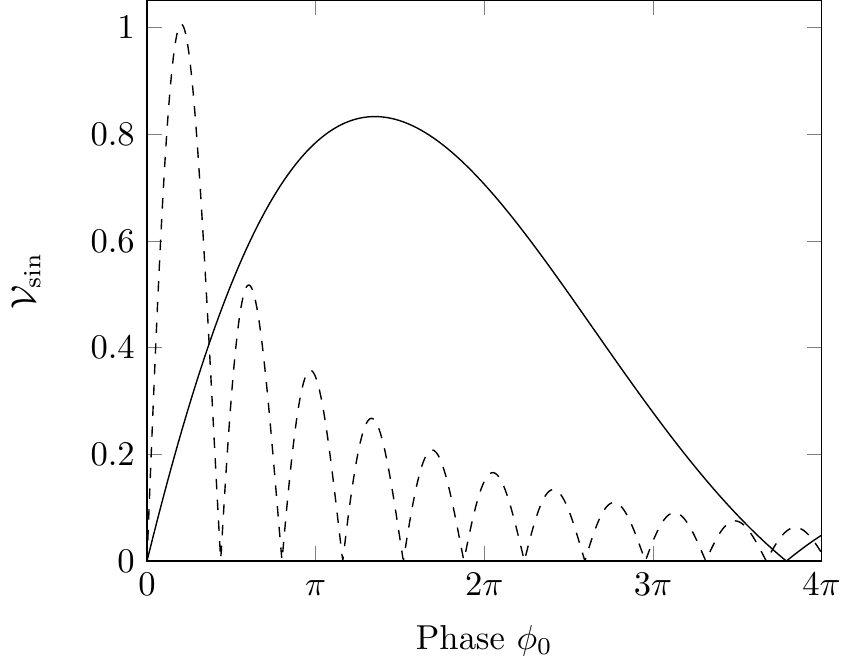}
  \caption{Expected quantum (solid) and classical (dashed) sinusoidal visibilities as a function of the phase-modulation parameter $\phi_0$.   This dependence on the pulse energy illustrates clearly the difference between the predictions.}
  \label{fig:visibility}
\end{figure}

The clear difference between the quantum and the classical pattern is captured by the sinusoidal fringe visibility, the ratio between the amplitude and the offset of a sine curve of period $D$ fitted to the density pattern (\ref{eq:pattern}),
\begin{equation}
 \V_{\sin} = 2 \left| B_1 \left[ \frac{t_1 t_2}{t_{\rm T} \left( t_1+t_2 \right)} \right] \right| \exp \left[ -\frac{2\pi^2 \sigma_x^2 t_2^2}{d^2 \left( t_1+t_2 \right)^2} \right]\,. \label{eq:VisSin}
\end{equation}
As shown in Fig.~\ref{fig:visibility}, the classical and the quantum prediction differ significantly: classical theory predicts many regions of low contrast as a function of $\phi_0$,  whereas the quantum prediction exhibits a slow $\phi_0$-dependence. 
The highest quantum visibility amounts to $83\,\%$ at $\phi_0=1.4\pi$.

\subsection{Accounting for decoherence}\label{sec:decoherence}
A realistic assessment of the proposed scheme must also include the influence of collisional and thermal decoherence \cite{hornberger2004theory}. This is incorporated into (\ref{eq:pattern})  by multiplying each Fourier component with a reduction factor of the form  
\begin{equation}
R_n = \exp \left\{ -\Gamma \left[ 1-f \left( \frac{nht_2}{mD} \right) \right] \left(t_1 + t_2 \right) \right\}, \label{eq:RedFact_dec}
\end{equation}
where $\Gamma$ gives the rate and $f(x)$ determines the spatial resolution of decoherence events of a certain class. In our simulation we accounted for collisions with residual gas particles, scattering and absorption of blackbody photons, and thermal emission of radiation using a  realistic microscopic description. Each process contributes another factor $R_n$ listed in the Supplementary Methods; 
the rate of thermal emission depends on time since the particle loses internal energy and cools during flight.

\subsection{Experimental constraints}
As a major concern for the successful implementation of the experiment, environmental decoherence must be kept sufficiently low. According to our simulations, collisional decoherence can be
essentially avoided at ultra-high vacuum pressures of $ 10^{-10}\,$mbar. 

Radiative decoherence is suppressed by choosing silicon spheres 
because they are essentially transparent at typical wavelengths of room temperature blackbody radiation. 
The thermal \emph{emission} of photons is determined by the internal temperature of the nanospheres, which is set in the trapping stage of the experiment.
A trapping intensity of $90\;\text{mW}/\upmu\text{m}^2$ leads to an initial heating rate $\partial_t T_\text{int}=200\;\text{K}/\text{s}$ and an equilibrium temperature of $1600\;\text{K}$. This high value is a consequence of the low blackbody emissivity of silicon \cite{palik1998handbook}, implying that the particle does not lose heat efficiently whilst in the trap.  
Nevertheless, due to the high refractive index  $n_{\rm Si}=3.48$ of silicon, the particle may be trapped for well in excess of a second before the temperature rises that high. This time corresponds to about $ 10^5$ trap oscillations, a sufficient period to perform parametric feedback cooling of the motion to $T=20\;\text{mK}$; see Supplementary Note 5. 

The low emissivity of silicon is the essential advantage compared to other materials such as silica, for which much work in this field has been done 
\cite{chang2010cavity,romero-isart2010toward,kiesel2013cavity}.
We find that to perform this experiment with silica would require cryogenic cooling of both apparatus and nanoparticle to $100\;$K, whereas thermal decoherence of silicon 
becomes important only at internal temperatures in excess of $1000\,$K 
Moreover, the high refractive index of silicon compared to the value $n_{\rm SiO2}=1.44$ of silica means than less optical power is required to trap the sphere and to monitor its position \cite{asenbaum2013cavity}.

As an additional advantage, silicon absorbs strongly at optical frequencies, which simplifies the detection of the interferogram. In principle, this would also affect the interaction with the grating laser, since a particle at the anti-node of the grating absorbs on average $n_0 = 0.12 \phi_0$ photons.
For a grating laser waist of $30\;$mm we anticipate a phase modulation  of $\phi_0/E_{\rm G}= 50\;\text{rad}/\text{mJ}$ and hence we can access $\phi_0 \leq 4\pi$.
The finite absorption of grating photons, which is included in the simulations, disturbs the interferogram little.

\section{Discussion}
We presented a viable scheme for high-mass nanoparticle interferometry, which employs only a single optical diffraction element and requires only moderate motional cooling. The setup would operate in ultra-high vacuum at room temperature. It is limited to masses up to $10^6\,$amu mainly by the growing Talbot time and free-fall distance \cite{nimmrichter2011testing}. Interferometry  in a microgravity environment could pave the way to even higher masses \cite{kaltenbaek2012macroscopic}.

Remarkably, with path separations of up to $150\,$nm and interrogation times of $300\,$ms, the presented scheme is  already sensitive to alternative theories beyond the Schr\"odinger equation. The renowned collapse model of continuous spontaneous localization (CSL) \cite{ghirardi1990markov} could be probed in its current formulation \cite{bassi2013models}. In fact, a successful demonstration of interference with a visibility exceeding
42\,\% would bound the localization rate to $\lambda_{\rm CSL} < 1.4\times 10^{-11}\,$Hz, a value at the lower end of recent estimates for this parameter \cite{adler2007lower,bassi2010breaking}; see Supplementary Discussion. Such a superposition experiment can be associated with a macroscopicity value of $\mu=18$ \cite{nimmrichter2013macroscopicity}, substantially exceeding that of every present-day matter-wave experiment and comparing well with the most ambitious micromirror superposition proposals \cite{marshall2003towards}.

\vspace{-2mm}
\section{Acknowledgements}
Funding by the EPSRC (EP/J014664/1), the Foundational Questions Institute (FQXi) through a Large Grant, and by the John F Templeton foundation (grant 39530) is gratefully acknowledged. The work was also partially supported by the European Commission within NANOQUESTFIT (No. 304886).

\vspace{-2mm}
\section{Author contributions}
HU concieved the interferometer and initiated the research. JB and SN designed the scheme and performed calculations and simulations. KH advised on the theory. All authors discussed the results and wrote the manuscript.


\newpage
\appendix
\onecolumngrid

\renewcommand{\theequation}{Supplementary Equation \arabic{equation}}
\renewcommand{\thefigure}{Supplementary Figure \arabic{figure}}
\renewcommand{\figurename}{\hspace{-0.6ex}}

\newcommand{\diff}{\text{\rm d}}

\newcommand{\vE}{\boldsymbol{E}}
\newcommand{\vn}{\boldsymbol{n}}

\newcommand{\cL}{ {\cal L} }

\newcommand{\Op}[1]{\mathsf #1} 
\newcommand{\oH}{ \Op{H} } 
\newcommand{\op}{ \Op{p} } 
\newcommand{\ox}{ \Op{x} } 

\renewcommand{\erf}{\text{\rm erf}}
\newcommand{\ih}{\frac{i}{\hbar}}

\newcommand{\qerw}[1]{\left\langle #1 \right\rangle}

\section{Supplementary Figures}

\begin{figure}[h!]
\includegraphics{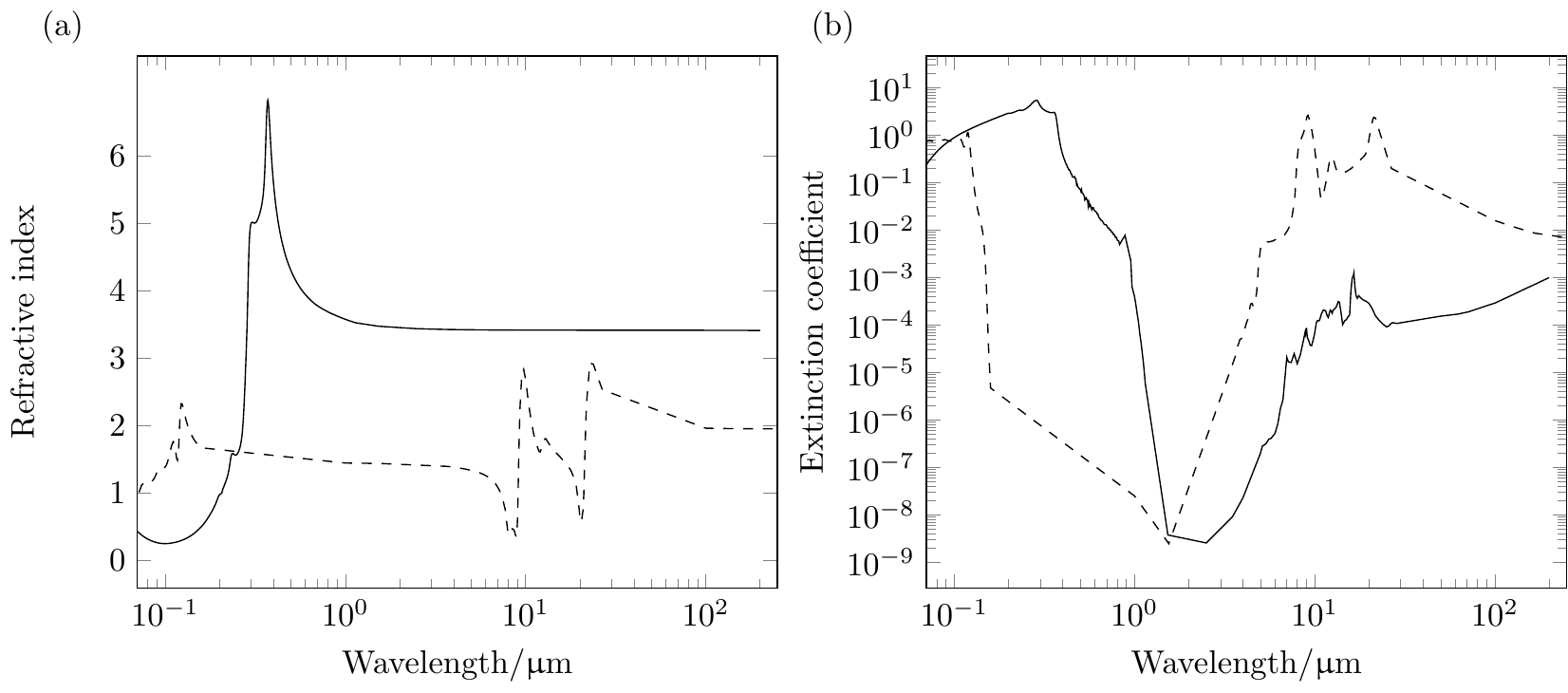}
\caption{\textbf{Spectral properties of silicon and silica.} Refractive indices (a) and absorption coefficients (b) for silicon (solid) and glass silica (dashed) as a function of the optical wavelength.
Data are for bulk materials as found in Ref.~\cite{palik1985handbook}. At wavelengths where absorption data is available and refractive index measurements are absent, the latter, which varies slowly, is found by linear interpolation.
Due to lack of tabulated data near the absorption minima, we have included values from recent absorption measurements \cite{lee2012ultra,steinlechner2013optical}.
The depicted spectra cover all relevant blackbody wavelengths for temperatures of the order of $10^3\,$K. We determine the static value of the dielectric function, as used to estimate the effect of collisions with background gas, by the refractive index value at the longest available wavelength.
}
\label{fig:Spectra}
\end{figure}

\begin{figure}
  \centering
  \includegraphics{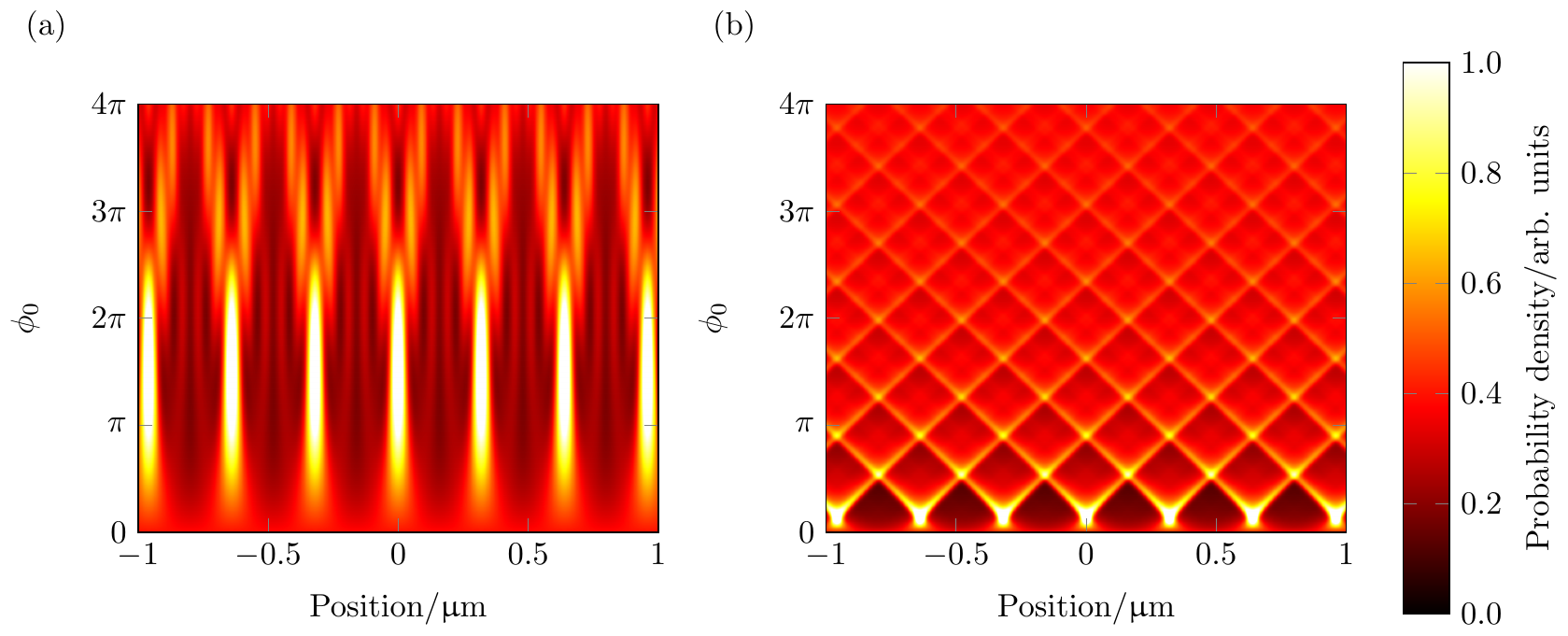}
  \caption{\textbf{Quantum and classical fringe patterns.} (a) Quantum and (b) classical fringe patterns for a grating period $d=355\;\text{nm}/2$ with $t_2=1.6t_T$ as we vary the maximum phase modulation $\phi_0$. The initial particle localization is $\sigma_x=10\,$nm and the free-flight time before the grating is $t_1=2t_T$ with $t_T=80\,$ms.}
  \label{fig:carpetVaryPhase}
\end{figure}

\begin{figure}
  \centering
  \includegraphics{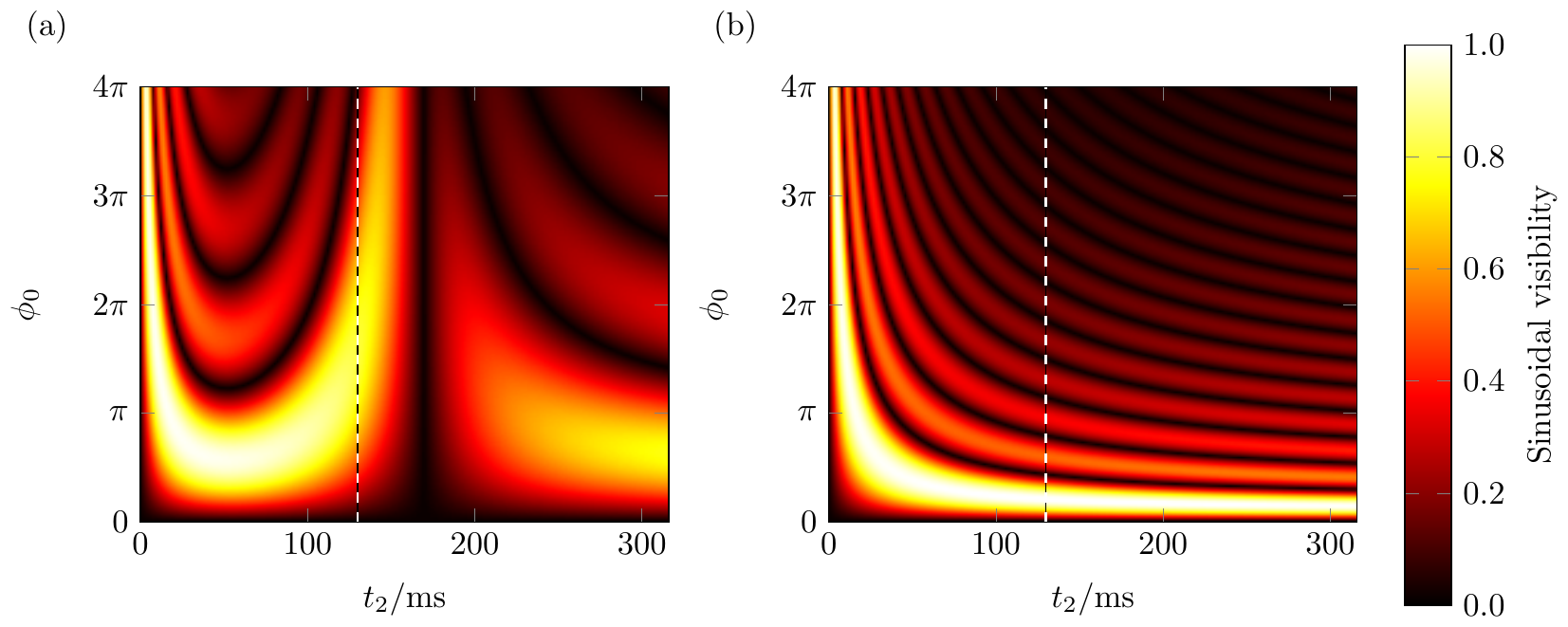}
  \caption{\textbf{Quantum and classical sinusoidal visibilities.} (a) Quantum and (b) classical sinusoidal visibilities for the parameters mentioned in the text, with varying grating phase $\phi_0$ and time $t_2$, for fixed $t_1=2t_T$. The vertical line shows the fixed time $t_2=1.6t_T$ for which we plot the visibilities as a function of phase in Fig.~3.}
  \label{fig:visibilitySurface}
\end{figure}

\begin{figure}
\includegraphics{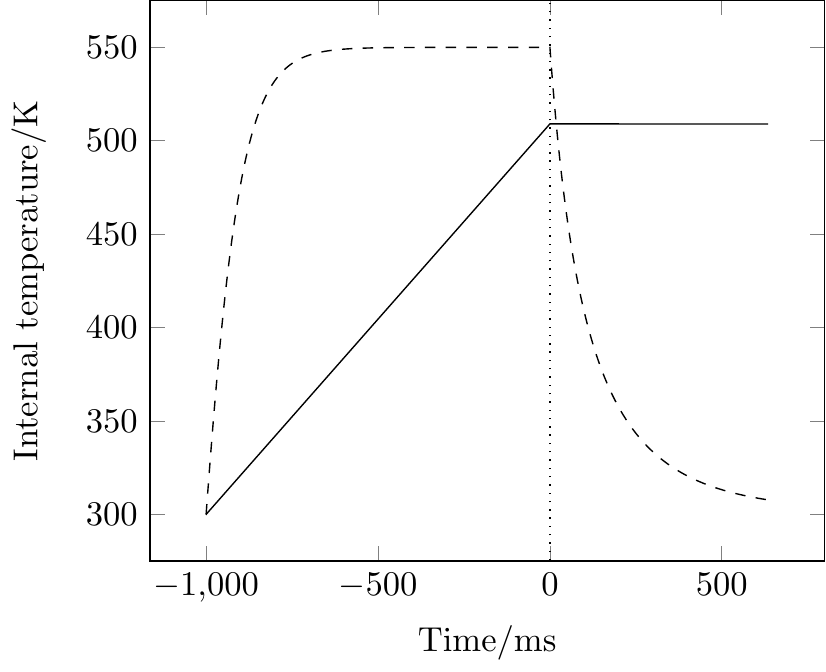}
\caption{\textbf{Internal heating and cooling of silicon and silica nanospheres.} Silicon (solid) and silica (dashed) nanospheres of $10^6\,$amu are exposed in the interval $(-1000\,{\rm ms}, 0)$ to light at wavelength $1550\,$nm focused with a 0.8 NA lens with intensity $I=90\,\text{mW}/\upmu\text{m}^2$ (silicon), $I=300\,\text{mW}/\upmu\text{m}^2$ (silica) chosen to yield a trap frequency $\nu_{\rm M}=200\,{\rm kHz}$. 
The silica particle reaches equilibrium temperature while the silicon particle undergoes constant heating, with an estimated equilibrium temperature of 1600\,K.
The enhanced infrared emissivity of silica gives rise to significant cooling after release, whereas silicon is almost perfectly isolated.
}
\label{fig:heating}
\end{figure}

\begin{figure}
\includegraphics{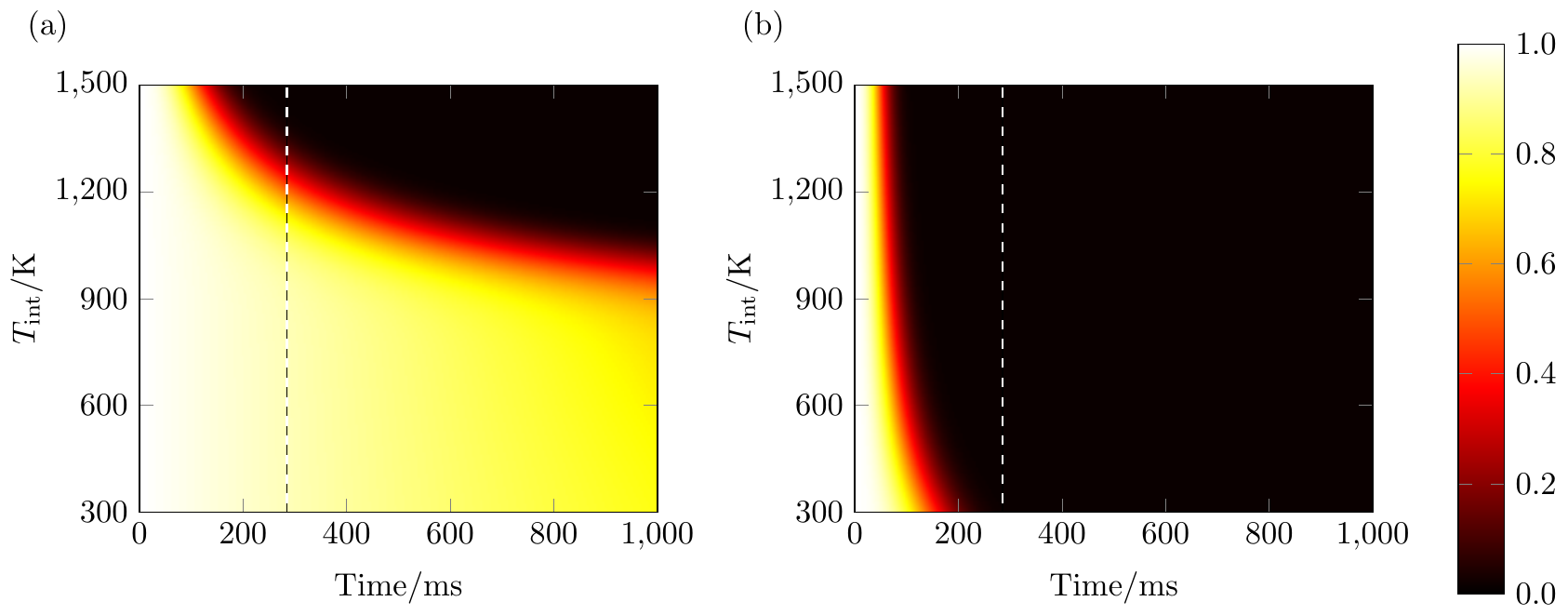}
\caption{\textbf{Reduction in visibility from decoherence for silicon and silica.} Reduction of the sinusoidal interference fringe visibility due to decoherence as a function of initial internal temperature and total time $t=t_1+t_2$, with the ratio $t_1/t_2=2/1.6$ as in our proposed experiment, for (a) silicon and (b) silica nanospheres. We assume a fixed environmental temperature of $T_\text{env}=300$\,K and a gas pressure of $p_g=10^{-10}\,$mbar.
The vertical dashed line corresponds to $t_1+t_2=3.6t_T$, which is the duration of the proposed interference experiment. We see clearly that silica exhibits significant decoherence due to thermal emission.
Decoherence due to collisions with background gas is only significant for times after approximately 500\,ms.}
\label{fig:decoherence}
\end{figure}

\section{Supplementary Notes}
\subsection*{Supplementary Note 1: Dielectric properties of nanospheres}
We summarize the important aspects of the linear interaction between spherical nanoparticles and electromagnetic fields. Given an electromagnetic wave of frequency $\omega=2\pi c/\lambda$, the linear response of a dielectric nanosphere of radius $R\ll\lambda$ is described by the complex polarizability, or susceptibility \cite{kreibig1995optical},
\begin{equation}
 \alpha = 4\pi\eps_0 R^3 \frac{\eps\left( \omega \right) - 1}{\eps\left( \omega \right) + 2}. \label{eq:susceptibility}
\end{equation} 
The relative permittivity is given by the square of the complex refractive index of the sphere material, $\eps=n^2$. We consider spheres which consist of many atoms, using refractive index data of the bulk material. 

The real part of the susceptibility determines the optical potential and dipole force acting on the particle in the presence of the standing-wave field $\vE \left( x,t \right)$,
\begin{equation}
 V\left( x,t \right) = -\frac{1}{4}\re \left\{ \alpha  \right\} \left| \vE \left( x,t \right) \right|^2 = -\frac{2\pi R^3 }{c} I\left( x,t \right) \re \left\{ \frac{\eps\left( \omega \right) - 1}{\eps\left( \omega \right) + 2} \right\}. \label{eq:Vgrating}
\end{equation} 
For a laser pulse of input power $P_{\rm G} \left( t \right)$ and spot area $a_{\rm G}$ that is reflected off a mirror to form a standing wave, we find the intensity $I\left( x,t \right) = 4P_{\rm G}\left( t \right) \cos^2 \left( 2\pi x/\lambda \right)/a_{\rm G}$ in the center of the spot. The optical dipole potential is the basis of the phase grating effect of such a standing-wave pulse, as the potential (\ref{eq:Vgrating}) gives rise to the eikonal phase 
\begin{equation}
 \phi \left( x \right) = -\frac{1}{\hbar} \int \diff t \, V\left( x,t \right) = \frac{2\re\left\{ \alpha \right\} E_{\rm G}}{\hbar c \eps_0 a_{\rm G}} \cos^2 \left( \frac{2\pi x}{\lambda} \right) \equiv \phi_0 \cos^2 \left( \frac{2\pi x}{\lambda} \right), \label{eq:phi_eik}
\end{equation} 
with $E_{\rm G} = \int \diff t\, P_{\rm G} \left( t \right)$ the pulse energy. The phase represents the classical action accumulated by a particle when illuminated by a standing-wave pulse in the Raman--Nath regime, where the laser pulse is sufficiently short (and weak) and where the particle is approximately at standstill \cite{berman1996atom,nimmrichter2008theory}. This periodic phase modulation of the motional quantum state of the particle gives rise to matter-wave interference, and it is the defining property of a phase grating. 

The imaginary part of the susceptibility (\ref{eq:susceptibility}) determines the light absorption power $P_{\rm abs} \left( x,t \right)=\sigma_{\rm abs} I\left( x,t \right)$, with the absorption cross section $\sigma_{\rm abs} = \omega \im\left\{ \alpha  \right\}/c\eps_0$. The average number of photons absorbed from the standing-wave pulse then reads as $n\left( x \right) = n_0\cos^2 \left( 2\pi x/\lambda \right) \equiv 2\beta\phi\left( x \right)$, introducing the material parameter
\begin{equation}
\beta = \frac{n_0}{2\phi_0} = \frac{\im \left\{ \alpha \right\}}{\re \left\{ \alpha \right\}} = \frac{3\,\im\left\{ \eps\left( \omega \right) \right\}}{\left| \eps\left( \omega \right) \right|^2+\re\left\{ \eps\left( \omega \right) \right\}-2}. \label{eq:beta}
\end{equation} 
Additional radiation pressure is related to the elastic dipole scattering of standing-wave photons into free space. The corresponding Rayleigh scattering cross-section, which is of higher order in $R/\lambda$ than the absorption, reads as $\sigma_{\rm sca} = \left( 2\pi/\lambda \right)^4 \left| \alpha \right|^2 / 6\pi \eps_0^2$. To quantify the number of scattered photons, $n_{\rm sca} \left( x \right) \equiv 2\eta \phi \left( x \right)$, we introduce the material- and size-dependent parameter 
\begin{equation}
 \eta = \left( \frac{2\pi}{\lambda} \right)^3 \frac{\left| \alpha \right|^2}{6\pi\eps_0 \re\left\{ \alpha \right\}} = \frac{2}{3} \left( \frac{2\pi R}{\lambda} \right)^3 \frac{\left| \eps\left( \omega \right) - 1 \right|^2}{\left| \eps\left( \omega \right) \right|^2+\re\left\{ \eps\left( \omega \right) \right\}-2} \label{eq:eta}.
\end{equation} 
Both absorption and Rayleigh scattering are incoherent additions to the conservative particle-light interaction given by (\ref{eq:Vgrating}) since they induce momentum diffusion. Hence, we require $\beta,\eta \ll 1$ for a standing-wave field to properly function as a pure phase grating \cite{hornberger2009theory,nimmrichter2011concept}. This is indeed the case for the choice of parameters considered in the main text. Using the spectral data for silicon and crystalline silica from \cite{palik1985handbook} 
at the grating wavelength $\lambda = \lambda_{\rm G} = 355\,$nm (see \ref{fig:Spectra}), we find $\beta = 0.06$ and $\eta = 6\times 10^{-4}$ for $m=10^6\,$amu.


The correct assessment of radiative decoherence and internal heating and cooling requires knowledge about the spectral absorption, scattering and emission rates at all relevant blackbody wavelengths. For this we use the refractive index data plotted in \ref{fig:Spectra}. Given a spectral rate function $\gamma\left( \omega \right)$, which is a dimensionless quantity, the total rate and power are obtained by integrating over all frequencies, $\Gamma = \int_0^\infty \diff\omega \, \gamma \left( \omega \right)$ and $P = \int_0^\infty \diff\omega \, \hbar\omega \gamma \left( \omega \right)$. 

For absorption and scattering, we assume that the radiation field is in thermal equilibrium at room temperature, $T_{\rm env} = 300\,$K. The absorption of radiation is then proportional to the absorption cross section and to the thermal occupation number, and by integrating over all directions of (isotropic) incident radiation we obtain
\begin{equation}
 \gamma_{{\rm abs}}\left(\omega\right)=\frac{\left(\omega/\pi c\right)^{2}\sigma_{{\rm abs}}\left(\omega\right)}{\exp\left(\hbar\omega/k_{B}T_{{\rm env}}\right)-1}=\frac{4\left(\omega R/c\right)^{3}/\pi}{\exp\left(\hbar\omega/k_{B}T_{{\rm env}}\right)-1}\im\left\{\frac{\eps\left(\omega\right)-1}{\eps\left(\omega\right)+2}\right\}. \label{eq:gabs}
\end{equation} 
A similar expression holds for the spectral scattering rate, 
\begin{equation}
 \gamma_{{\rm sca}}\left(\omega\right)=\frac{\left(\omega/\pi c\right)^{2}\sigma_{{\rm sca}}\left(\omega\right)}{\exp\left(\hbar\omega/k_{B}T_{{\rm env}}\right)-1}=\frac{8\left(\omega R/c\right)^{6}/3\pi}{\exp\left(\hbar\omega/k_{B}T_{{\rm env}}\right)-1}\left|\frac{\eps\left(\omega\right)-1}{\eps\left(\omega\right)+2}\right|^{2}. \label{eq:gsca}
\end{equation} 
In the case of thermal emission of radiation, we must once again use the absorption cross section. We do, however, allow for a different internal temperature $T_{\rm int}$ of the particle, which may not be in thermal equilibrium with the environment due to trap heating. For all our estimates we consider here the worst-case scenario, where the particle is much hotter than the environment, $T_{\rm int} \gg T_{\rm env}$. In this case there is no stimulated emission by the radiation background and the spectral emission rate is governed by the \textit{Boltzmann} factor \cite{hansen1998thermal},
\begin{equation}
\gamma_{{\rm emi}}\left(\omega,T_{{\rm int}}\right)=\left(\frac{\omega}{\pi c}\right)^{2}\sigma_{{\rm abs}}\left(\omega\right)\exp\left(-\frac{\hbar\omega}{k_{B}T_{{\rm int}}}\right)=\frac{4}{\pi}\left(\frac{\omega R}{c}\right)^{3}\exp\left(-\frac{\hbar\omega}{k_{B}T_{{\rm int}}}\right)\im\left\{\frac{\eps\left(\omega\right)-1}{\eps\left(\omega\right)+2}\right\}. \label{eq:gemi}
\end{equation} 
Additional corrections due to the finite heat capacitance of the particle are neglected. 
Notice the subtle difference between the Planck factor in (\ref{eq:gabs}) and the Boltzmann factor in (\ref{eq:gemi}), which makes a dramatic difference at low frequencies, $\hbar\omega \ll k_B T_{\rm int,env}$. Whereas the Planck factor gives rise to an enhanced absorption of these low-energy photons, the same photons are emitted from hot particles at a significantly reduced rate due to the lack of stimulated emission \cite{hansen1998thermal}. Although this difference has been overlooked in other nanosphere trapping proposals \cite{chang2010cavity,romero-isart2010toward}, we will see below that this effect can lead to considerable changes in the trap heating rates. 

The Boltzmann form (\ref{eq:gemi}) ceases to be valid when the internal temperature approaches thermal equilibrium with the environment; it then underestimates the emission rate and, thus, the radiative damping of internal energy. The particle heats up faster and cools down more slowly. Using the above expression can therefore be regarded as a conservative estimate for our reasoning concerning decoherence and heating.
\subsection*{Supplementary Note 2: Dipole trapping, particle size, and initial localization}

To achieve significant visibility, we require the position  uncertainty of the particle in the trap to be $\sigma_x\lesssim d/(2\pi)\approx 30\,\text{nm}$.
While challenging, this localization is feasible using parametric feedback to `cool' the center of mass motion of the particle.
Employing the method described by Gieseler et al. \cite{gieseler2012subkelvin}, under the paraxial approximation we find a sensitivity to positional changes of the particle about the center of the trap of
$8\alpha/(\epsilon_0w_0^3\lambda\sqrt{\pi})$
where $w_0$ is the waist of the laser spot in which the particle is trapped. The strong dependence upon this waist suggests that a high numerical aperture lens should be used for trapping. Using a numerical aperture of 0.8, one can expect a relative signal change of $2\times 10^{-8}\;\text{nm}^{-1}$.

The fractional change in power which one can resolve is ultimately limited by shot noise in the photon number.
However, increasing the laser power has the unwanted effect of increasing the trap frequency, and hence increases the bandwidth with which one must resolve this fractional change.
Fortunately, the mechanical oscillation can be  expected to be characterized by an extraordinarily high Q factor and, by employing boxcar averaging, one may increase the effective integration time.

We choose a mechanical trap frequency of $200\;\text{kHz}$ which is similar to previously demonstrated traps and, for silicon, requires a modest $53\,\text{mW}$. For this power, and for a photodiode responsivity of $1.0\;\text{A}/\text{W}$, we find a relative shot noise of $1.7\times 10^{-9}/\sqrt{\text{Hz}}$ and thus, using 100 periods for boxcar averaging, a position uncertainty $\sigma_x<10\;\text{nm}$ can be achieved.
The \emph{internal} heating of the particle from photon absorption, as discussed in \ref{sec:internalHeating}, places a limit on the time for which we may trap and implement feedback cooling.  For typical values, we anticipate about $ 10^5$ oscillations within which to perform this cooling.

Additionally, since the scattering force causes a size-dependent offset of the equilibrium position from the laser focus, the time-averaged position along the optical axis provides a direct measure of particle size.  The particle will be displaced by approximately 10\% of its radius and so, by integrating the error signal which we use for feedback over a few hundred milliseconds, we may discern the relative particle size with sub-nanometre precision. The displacement is far more significant at lower NA, to which the trap may be reduced transiently for the express purpose of determining particle size.

\subsection*{Supplementary Note 3: Position stability of the grating}
As noted in the main text, the position stability of the grating must be similar to the initial localization of the state $\sigma_x$.  The positions of the (anti-)nodes in the standing lightwave which forms this grating are fixed relative to the position of the mirror.  Achiving the necessary position stability of this mirror is challenging, but the recently demonstrated `OTIMA' experiment~\cite{haslinger2013universal}, which employs a similar laser to ionize macro-molecules, has shown that this stability can be achieved when one accounts correctly for the small absorption and consequent heating in the mirrors.

\subsection*{Supplementary Note 4: Position detection and experimentally accessible interferograms}
The interferogram is obtained by recording the arrival position of individual nanoparticles on a glass slide at a fixed distance below the source.
The position of a nanosphere must be resolved with positional accuracy exceeding $\mu d/3\sim 100\;\text{nm}$ which, given that one may interrogate individual nanoparticles for an essentially unlimited time, may be achieved by fitting the recorded image to the known point-spread function of the imaging system. For silicon, absorption imaging with visible light can be used. We note that an uncertainty in time $t_2$ accrues due to the uncertainty in the vertical component of the initial velocity.  For the distances in the proposed experiment, we find a relative uncertainty of $\sigma_{t_2}/t_2 \approx 0.5\%$, which is negligibly small.


The experimental apparatus enforces a fixed free-fall distance and limited range over which to vary $t_1$; hence, rather than directly accessing the fringe pattern as a function of $t_2$ for fixed $t_1$ as shown in Fig.~2 of the main text, we instead examine the spatial distribution as a function of the phase modulation parameter $\phi_0$, which may be varied by controlling the pulse energy of the grating laser.  The corresponding plot is shown in \ref{fig:carpetVaryPhase}.
When designing the experiment, one is free to choose the free-flight times $t_1$ and $t_2$.  For a fixed $t_1=2t_T$ we plot the expected sinusoidal visibilities, for both quantum and classical cases, as a function of $t_2$ and of $\phi_0$.  The resulting surfaces plots are shown in \ref{fig:visibilitySurface}.


\subsection*{Supplementary Note 5: Internal heating and cooling}\label{sec:internalHeating}

We estimate the internal heating of the nanosphere in the trap by solving a rate-balance equation for the internal energy $U \left( t \right)$ as a function of time. The energy increases by absorbing either laser photons or blackbody radiation at room temperature, and it decreases by emitting thermal radiation. Given the specific heat $c_m$ of the sphere material, we can identify $\diff U/\diff t = mc_m \diff T_{\rm int} / \diff t$ and write 
\begin{equation}
 m c_m \frac{\diff T_{\rm int}}{\diff t} = \frac{4\pi I_{\rm T} \omega_{\rm T} R^3}{c} \im\left\{\frac{\eps\left(\omega_{\rm T}\right)-1}{\eps\left(\omega_{\rm T}\right)+2}\right\} + \int \diff \omega \left[ \gamma_{\rm abs} \left( \omega \right) - \gamma_{\rm emi} \left( \omega,T_{\rm int} \right) \right] \hbar\omega, \label{eq:heating}
\end{equation} 
using the above spectral rate expressions for absorption (\ref{eq:gabs}) and emission (\ref{eq:gemi}) of thermal radiation. The trap laser intensity and laser frequency are denoted by $I_{\rm T}$ and $\omega_{\rm T} = 2\pi c/\lambda_{\rm T}$.


Here, we fix the mechanical trap frequency at $\nu_{\rm M} = 200\,$kHz, which requires trapping intensities of $90\,$mW/$\mu$m$^2$ and $300\,$mW/$\mu$m$^2$ for $10^6\,$amu silicon and silica nanospheres, respectively. Assuming a specific heat of $c_m=700\,$J/kg\,K for both materials, and using the interpolated refractive index data of \ref{fig:Spectra}, a numerical evaluation of (\ref{eq:heating}) yields the heating curves depicted in \ref{fig:heating}. Note that, due to the lack of absorption data at the trap laser wavelength $\lambda_{\rm T} = 1550\,$nm in Ref.~\cite{palik1985handbook}, we use separate values for the absorption measured in Refs.~\cite{lee2012ultra,steinlechner2013optical}. They amount to $\im\left\{ n_{\rm Si} \right\} \approx 3.7\times 10^{-9}$ and $\im\left\{ n_{\rm SiO2} \right\} \approx 2.5\times 10^{-9}$, and they are included in \ref{fig:Spectra}.

The striking difference in the heating curves of silicon and silica are due to two reasons: (i) The high refractive index of silicon, $\re\left\{ n_{\rm Si} \right\} = 3.48$ versus $\re\left\{ n_{\rm SiO2} \right\} = 1.44$, which implies  less required laser power, i.e.\ less absorption. (ii) The different absorption spectra in the thermally accessible regime of mid- to far-infrared wavelengths; there, silicon exhibits much less absorption than silica, which inhibits energy damping by emission.

The different thermal emission behavior becomes evident when the particle is released from the trap after a conservatively estimated trapping time of, say, $1000\,$ms. \ref{fig:heating} shows the internal temperature of the silicon and silica nanoparticles as a function of time after release. Silicon has a much lower emissivity at thermal infrared wavelengths than silica, which suppresses the emission damping completely over the time scale of the proposed interference experiment. Silica nanoparticles, on the other hand, lose a significant amount of their initial energy. This must be taken into account in the assessment of decoherence by thermal emission of radiation.

\section{Supplementary Discussion}

\subsection*{State reduction by continuous spontaneous localization (CSL)}

The proposed high-mass interferometer scheme can be used to test the predictions of certain macrorealistic collapse models, which were conceived in order to induce a breakdown of the quantum superposition principle and thereby reconcile quantum with classical mechanics at the macroscale \cite{bassi2013models}. The best studied one is the theory of continuous spontaneous localization (CSL) \cite{ghirardi1990markov}. The master equation describing the effect of CSL on the center-of-mass motion of a nanoparticle of mass $m$ predicts a fringe reduction equivalent to a decoherence process with the parameters \cite{vacchini2007precise,nimmrichter2011testing}
\begin{equation}
 \Gamma_{\rm CSL} = \left( \frac{m}{\rm 1\,amu} \right)^2 \lambda_{\rm CSL}, \quad g\left( x \right) = \exp\left( -\frac{x^2}{4r_c^2} \right), \quad f\left( x \right) = \frac{\sqrt{\pi}r_c}{x}\erf\left( \frac{x}{2r_c} \right).
\end{equation} 
Here, the CSL localization length is conventionally set to $r_c=100\,$nm, and the free rate parameter is currently estimated to be in the range of $\lambda_{\rm CSL} \sim 10^{-10 \pm 2}\,$Hz \cite{adler2007lower,bassi2010breaking}. 

If the CSL effect existed, the sinusoidal fringe visibility 
Eq.~(5) would be reduced by the factor $R_1^{\rm CSL}$, as obtained by plugging the above CSL parameters into  
Eq.~(6). That is to say, if one would measure at least half of the expected unmodified visibility after $t_1+t_2=3.6t_T=284\,$ms in the proposed setup, the CSL rate parameter would be bounded by $\lambda_{\rm CSL} < 1.4\times 10^{-11}\,$Hz.

\section{Supplementary Methods}
\subsection{Phase-space description of the interference effect}

Next we give a detailed theoretical description of the near-field interference effect discussed in the main text. The model includes the influence of environmental decoherence, an external acceleration, as well as incoherent additions to the grating interaction due to absorption and scattering. The latter are shown to be negligible for the materials considered in the main text, but they may become relevant in other cases. 

We treat the problem in one dimension along the grating axis $x$,
where the interference pattern builds up. This assumes that the
motional state of the particles along $y,z$ remains separable from the $x$-motion
at all times. In particular, we assume that the grating interaction
and the detection do not depend on the $y,z$-coordinates. In practice, this requires a properly aligned trap (its principal axis parallel to $x$) and a wide grating laser spot, which guarantees a uniform illumination
of the particles irrespectively of their $y,z$-position.

\subsubsection{Characteristic function representation}

We use the characteristic function representation to
conveniently include all decoherence effects and to be able to compare
the quantum and the classical expectation within the same formalism. 
Given the Wigner function $w\left(x,p\right)$ of the one-dimensional motional
state of the particle \cite{schleich2001quantum}, we define the
characteristic function as the Fourier transform
\begin{equation}
\chi\left(s,q\right)=\int\diff x\diff p\, w\left(x,p\right)e^{i\left(qx-ps\right)/\hbar}=\tr{\rho\exp\left[\ih\left(q\ox-\op s\right)\right]},
\end{equation}
with $\ox,\op$ the position and momentum operators. The classical
analogue, where particles move on ballistic trajectories, is obtained by replacing the Wigner function with the classical phase-space distribution function $f_{\rm cl}\left(x,p\right)$.

\subsubsection{Initial state}

We start from the motional state of the particle when it is released from the optical trap. To a good approximation, this initial state can be modeled by a thermal harmonic oscillator state. Given the trap frequency $\nu_{\rm M}$ and the motional temperature
$T$, the state is represented by the Gaussian Wigner function 
\begin{equation}
w_0\left(x,p\right)=\frac{1}{2\pi\sigma_{x}\sigma_{p}}\exp\left(-\frac{x^{2}}{2\sigma_{x}^{2}}-\frac{p^{2}}{2\sigma_{p}^{2}}\right), \quad \chi_0\left(s,q\right)=\exp\left(-\frac{\sigma_{x}^{2}q^{2}+\sigma_{p}^{2}s^{2}}{2\hbar^{2}}\right),\label{eq:chi_thermaltrap}
\end{equation}
where the standard deviations in position and momentum read as \cite{schleich2001quantum}
\begin{equation}
\sigma_{x}=\sqrt{\frac{\hbar}{4\pi m \nu_{\rm M}}\coth\left(\frac{h \nu_{\rm M}}{2k_{B}T}\right)},\qquad \sigma_{p}=\sqrt{\pi \hbar m \nu_{\rm M} \coth\left(\frac{h \nu_{\rm M}}{2k_{B}T}\right)}.
\end{equation}
For the realistic values considered here, $\nu_{\rm M} = 200\,$kHz and $T = 20\,$mK, the Wigner function in \eqref{eq:chi_thermaltrap} is practically indistinguishable from the classical thermal phase-space distribution of the harmonic oscillator, since $h \nu_{\rm M} \ll k_{B}T$ and thus $\coth x\approx1/x$.

On the other hand, if we compare the momentum spread to the elementary momentum unit $h/d$ of the standing-wave grating with $2d=355\,$nm, we find $\sigma_p d/h \approx 10^4$. That is, the initial trap state extends over many grating momenta and the time-evolved Gaussian envelope will extend over similarly many grating periods. As long as we are interested in the central part of the interference pattern, where the envelope is flat, it is therefore 
justified to use the limiting expression
\begin{equation}
 \chi_0\left(s,q\right) \approx \frac{\sqrt{2\pi} \hbar}{\sigma_p} \exp\left(-\frac{\sigma_{x}^{2}q^{2}}{2\hbar^{2}}\right) \delta \left( s \right)\,,\label{eq:chi_trap_delta}
\end{equation} 
a standard approximation that simplifies the calculation considerably and that is also employed in other near-field interference schemes \cite{hornberger2004theory,juffmann2012new,nimmrichter2011concept}. Finite-size fringe effects can be disregarded provided the detection window is reasonably small compared to the spread of the final fringe pattern.

\subsubsection{Free propagation in the presence of acceleration and decoherence}

Between the trap , the grating pulse, and the final detection the particle state evolves freely for the times $t_1$ and $t_2$. Ideally, this is represented by a shearing transformation of the form $\chi_t ( s,q ) = \chi_0 \left( s-qt/m,q \right)$ in phase space \cite{schleich2001quantum}. A more realistic assessment must include the influence of possible environmental disturbances as well as inertial forces. 

Here, we account for the presence of a (time-dependent) external acceleration $a\left( t \right)$, which may be a small component of gravity in the case of grating misalignment, for instance. This yields the Hamiltonian $\oH( t ) = \op^2 /2m + m a ( t) \ox$ for the one-dimensional motion along the $x$-axis. The time evolution of the particle's quantum state, $\partial_t \rho = -i[\oH\left( t \right),\rho]/\hbar + \cL\left( t \right) \rho$, shall include also decoherence effects represented by the (time-dependent) generator $\cL \left( t \right)$. The latter describes the random unitary state transformation associated to each independent decoherence process: gas collisions, Rayleigh scattering of blackbody
photons, and thermal emission or absorption of radiation. 
Each of these contributes a generator of the form \cite{hornberger2004theory}
\begin{equation}
\la x|\cL \left( t \right) \rho|x'\ra=\Gamma\left(t\right)\left[g\left(x-x'\right)-1\right]\la x|\rho|x'\ra,\label{eq:decoherenceProcess}
\end{equation}
in position representation. It describes random jump events $\la x|\rho|x'\ra\to g \left(x-x'\right)\la x|\rho|x'\ra$
at (possibly time-dependent) rates $\Gamma\left(t\right)$. The decoherence function is normalized
to $g\left(0\right)=1$, and its characteristic width represents
the finite resolution of the decoherence effect. Only those superpositions
with a path separation $x-x'$, for which $g$ vanishes approximately,
decohere at the full rate $\Gamma\left(t\right)$.

In phase space, Eq.~(\ref{eq:decoherenceProcess}) is represented by a momentum averaging transformation, which can also be understood in a classical  picture.
The characteristic function transforms in both the quantum and the classical case as 
\begin{equation}
\chi_{t}\left(s,q\right)=\chi_{0}\left(s-\frac{qt}{m},q\right)\exp\left\{ \ih\left[q\Delta x\left( t \right)-\Delta p\left( t \right)s\right]+\int_{0}^{t}\diff\tau\Gamma\left(\tau\right)\left[g\left(s-\frac{q\tau}{m}\right)-1\right]\right\}. \label{eq:chi_prop}
\end{equation}
Here, we introduced the momentum and position shifts $\Delta p\left( t \right) = m\int_0^t \diff t' \, a\left( t' \right)$ and $\Delta x\left( t \right) = \int_0^t \diff t'\, \Delta p\left( t' \right)/m$.

\subsubsection{Grating transformation}

The key difference between the quantum and the classical 
description of the interferometer is the state transformation at the standing-wave grating. 
The quantum transformation describes a modulation $\phi\left( x \right)$ of the matter-wave phase by the term (\ref{eq:phi_eik}), which leads to interference after the grating.
The classical transformation, on the other hand, is given by the accumulated momentum kick $Q\left(x\right)=\hbar\partial_{x}\phi\left(x\right)$
exerted on the particle by the dipole force in the standing wave, which leads to a periodic lensing effect after the grating \cite{hornberger2009theory}.

In phase space, the grating transformation is a convolution, conveniently expressed in terms of the \emph{Talbot}
coefficients $B_{n}\left(\xi\right)$ \cite{hornberger2004theory,nimmrichter2008theory,hornberger2009theory},
\begin{equation}
\chi\left(s,q\right)\to \sum_{n}B_{n}\left(\frac{s}{d}\right) \chi\left(s,q + n\frac{h}{d}\right).\label{eq:gratingtrafo}
\end{equation}
Here, $d=\lambda_{\rm G}/2$ denotes the grating period determined by the laser wavelength $\lambda_{\rm G}$. The Talbot coefficients of a standing-wave phase grating assume a simple analytical form in terms of Bessel functions \cite{hornberger2009theory}, $B_{n}\left(\xi\right)=J_{n}\left(\phi_{0}\sin\pi\xi\right)$. 
The corresponding classical transformation can be brought
into the same form \eqref{eq:gratingtrafo}, substituting $B_{n}\left(\xi\right)$
by their classical counterparts $C_{n}\left(\xi\right)=J_{n}\left(\phi_{0}\pi\xi\right)$.

Apart from the coherent phase modulation described by the above grating transformation, random photon absorption and scattering 
events must also be taken into account. They lead to the following modification of the Talbot coefficients. 

Photon absorption transfers momentum in units of the photon
momentum $h/2d$ to the particle. This stochastic absorption process
is described by the standard master equation \cite{nimmrichter2010master},
\begin{equation}
\cL_{{\rm abs}}\rho=\gamma_{{\rm abs}}\left[\cos\left(\frac{\pi\ox}{d}\right)\rho\cos\left(\frac{\pi\ox}{d}\right)-\frac{1}{2}\left\{ \cos^{2}\left(\frac{\pi\ox}{d}\right),\rho\right\} \right],\label{eq:gratingAbsME}
\end{equation}
which must be added to the coherent time evolution due to the standing-wave field. The absorption rate $\gamma_{{\rm abs}}$
determines the mean number of absorbed photons, $n_{0} = 2\beta\phi_0$,
with the parameter $\beta$ defined in (\ref{eq:beta}) \cite{nimmrichter2011concept}.

The master equation  $\partial_{t}\rho=\cL_{{\rm abs}}\rho$
can be integrated explicitly by neglecting the motion during the pulse. 
The resulting transformation is represented by a convolution in phase space similar to \eqref{eq:gratingtrafo}, but the Fourier coefficients are now
given by modified Bessel functions, 
\begin{equation}
R_{n}^{({\rm abs})}\left(\xi\right)=\exp\left(-n_{0}\frac{1-\cos\pi\xi}{2}\right)I_{n}\left(n_{0}\frac{1-\cos\pi\xi}{2}\right).
\end{equation}
Applying both the coherent and the absorption transformation subsequently, we arrive
at the overall grating transformation. It is obtained by replacing
the coherent Talbot coefficients with
\begin{equation}
B_{n}\left(\xi\right)=e^{-\zeta_{{\rm abs}}\left(\xi\right)}\left[\frac{\zeta_{{\rm coh}}\left(\xi\right)+\zeta_{{\rm abs}}\left(\xi\right)}{\zeta_{{\rm coh}}\left(\xi\right)-\zeta_{{\rm abs}}\left(\xi\right)}\right]^{n/2}J_{n}\left[\sgn\left\{ \zeta_{{\rm coh}}\left(\xi\right)-\zeta_{{\rm abs}}\left(\xi\right)\right\} \sqrt{\zeta_{{\rm coh}}^{2}\left(\xi\right)-\zeta_{{\rm abs}}^{2}\left(\xi\right)}\right],
\end{equation}
where we abbreviate $\zeta_{{\rm coh}}\left(\xi\right)=\phi_{0}\sin\pi\xi$
and $\zeta_{{\rm abs}}\left(\xi\right)=n_{0}\sin^{2}\left(\pi\xi/2\right)=\beta\phi_{0}\left(1-\cos\pi\xi\right)$.
The result follows from several steps of calculation using Graf's
addition theorem for Bessel functions \cite{abramowitz1965handbook}. Note
that the present form differs from the result given in \cite{hornberger2009theory},
where the absorption effect was described in terms of a classical
random walk model. The classical analogue, where the coherent phase
modulation is replaced by a position-dependent momentum kick, is obtained by replacing $\zeta_{{\rm coh}}\left(\xi\right)$ with $\phi_{0}\pi\xi$.

Should Rayleigh scattering be of any concern, one could implement
an additional master equation describing the absorption and the subsequent
re-emission of laser photons in the form of dipole radiation. A tedious
but straightforward calculation would then lead to the modification
$B_{n}\left(\xi\right)\to\sum_{j}B_{n-j}\left(\xi\right)R_{j}\left(\xi\right)$,
with\begin{equation}
R_{n}^{\rm (sca)}\left(\xi\right)=\exp\left[-\frac{n_{R}}{2}\left(1-3\cos\pi\xi\frac{\sin\pi\xi-j_{1}\left(\pi\xi\right)}{2\pi\xi}\right)\right]I_{n}\left[\frac{n_{R}}{2}\left(3\frac{\sin\pi\xi-j_{1}\left(\pi\xi\right)}{2\pi\xi}-\cos\pi\xi\right)\right].\end{equation}
The mean number of scattered photons, $n_R = 2\eta \phi_0$, is proportional to the material parameter (\ref{eq:eta}). 

For the materials considered here, we find $\beta,\eta \ll 1$, and so we neglect the absorption and scattering effect in the main text (although they are included in the simulation). This may no more be the case for other materials or greater masses.

\subsubsection{The interference effect}

The interference scheme can now be assessed in phase space using all the above ingredients. The approximate initial state (\ref{eq:chi_trap_delta}) is evolved for the time $t_1$ after release, using Eq.~(\ref{eq:chi_prop}). Then the grating pulse is applied and the state is propagated for another time $t_2$. This results in the characteristic function 
\begin{eqnarray}
 \chi \left( s,q \right) &=& \sum_{n}B_{n}\left(\frac{s}{d}-\frac{qt_2}{md}\right) \exp\left\{ \ih\left[q\Delta x\left( t_2 \right)-\Delta p\left( t_2 \right)s\right]+\int_{0}^{t_2}\diff\tau\,\Gamma\left(t_1+\tau\right)\left[g\left(s-\frac{q\tau}{m}\right)-1\right]\right\} \nonumber \\
 && \times \exp\left\{ \ih\left[\left( q + n\frac{h}{d} \right) \Delta x\left( t_1 \right) - \Delta p\left( t_1 \right)\left( s - \frac{qt_2}{m} \right) \right]+\int_{0}^{t_1}\diff\tau\,\Gamma\left(\tau\right)\left[g\left(s-q\frac{t_2+\tau}{m}-n\frac{h\tau}{md}\right)-1\right]\right\} \nonumber \\
 && \times \chi_{0}\left(s- q \frac{t_1 + t_2}{m} - n\frac{ht_1}{md},q + n\frac{h}{d} \right).
\end{eqnarray} 
The density distribution of the particle state, i.e. the interferogram, is obtained by evaluating the integral
\begin{eqnarray}
w\left(x\right) &=& \la x|\rho|x\ra=\frac{1}{2\pi\hbar}\int\diff q\,\chi_{3}\left(0,q\right)\exp\left(-\frac{iqx}{\hbar}\right) \nonumber \\
 &=& \frac{m}{\sqrt{2\pi}\sigma_p \left( t_1 + t_2 \right)} \sum_{n} \exp\left[\frac{2\pi i n}{\mu d} \left( x - \delta x \right) \right] B_{n}\left(\frac{n t_2}{\mu t_T}\right) \exp\left\{ -\frac{1}{2} \left[ \frac{2\pi n \sigma_x t_2 }{d \left( t_1 + t_2 \right)} \right]^2 \right\} \nonumber \\
 && \times \exp\left\{ \int_{0}^{t_2}\diff\tau\,\Gamma\left(t_1+\tau\right)\left[g\left(\frac{nh\tau}{\mu m d}\right)-1\right] + \int_{0}^{t_1}\diff\tau\,\Gamma\left(\tau\right)\left[g\left(\frac{nht_2}{\mu m d}\, \frac{t_1 - \tau}{t_1}\right)-1\right]\right\}.\nonumber\\ \label{eq:w_final}
\end{eqnarray}
The Talbot time $t_{T}=md^{2}/h$ appears as the natural time scale. We are left with a periodic fringe pattern, where the particle density oscillates at the geometrically magnified grating period $D = \mu d$, with $\mu = \left( t_1+t_2 \right)/t_1$. In the presence of a time-dependent homogeneous acceleration $a\left( t \right)$ the fringe pattern is shifted by
\begin{equation}
 \delta x = \Delta x\left( t_1+t_2 \right) - \mu \Delta x\left( t_1 \right) = \int_0^{t_1+t_2}\diff t \int_0^t \diff \tau \, a\left( \tau \right) - \mu \int_0^{t_1}\diff t \int_0^t \diff \tau \, a\left( \tau \right)\,.
\end{equation} 
In the main text, 
Eq.~(4), we have given the result in the absence of decoherence and acceleration.

In the limit of a perfect point source, $\sigma_{x}\to0$, the result
is a Talbot-like image of the grating \cite{brezger2003concepts},
whereas the fringe amplitudes are exponentially reduced for finite
$\sigma_{x}$. The fringe pattern washes out completely once $\sigma_{x}\gtrsim\mu dt_{1}/2\pi t_{2}$.
The full Talbot effect, i.e. the reconstruction of the grating mask
profile (albeit magnified by the factor $\mu$), is observed when
$t_{2}/\mu$ is an integer multiple of the Talbot time. In the 
case of a pure phase grating, however, this reconstruction is a flat
line, since $B_{n}\left(N\in\mathbb{N}\right)=\delta_{n,0}$. 
The classical case is obtained by replacing the Talbot coefficients with their classical counterparts. 

Regions of high contrast lie in between the Talbot orders, around
$t_{2}/\mu t_{T}\approx\left(2N+1\right)/2$. At small times $t_{2}<t_{T}$
the quantum and the classical fringe pattern are hard to distinguish.
Long flight times, on the other hand, increase the deteriorating influence
of decoherence and source averaging. One should therefore generally work in
the regime $t_{1,2}\approx t_{T}$.

Note that the strictly periodic expression (\ref{eq:w_final}) for the particle density distribution is only valid for measurements in the center of the Gaussian envelope of the dispersed particle state. The average particle detection probability in a detection window of width 
$W$ at the center is given by 
\begin{equation}
P_{\rm det} = \qerw{w\left(x\right)}_{W}\approx\frac{Wm}{\sqrt{2\pi}\sigma_{p}\left(t_{1}+t_{2}\right)}.
\end{equation}

\subsection{Influence of decoherence}

We now incorporate the four relevant environmental decoherence effects into the calculation. The last line of the expression (\ref{eq:w_final}) for the predicted fringe pattern represents the effect of a generic decoherence process of the type (\ref{eq:decoherenceProcess}): it multiplies the fringe amplitudes by the reduction factors
\begin{equation}
 R_n = \exp\left\{ \int_{0}^{1}\diff\vartheta\,\left[ t_2 \Gamma\left(t_1+t_2 \vartheta\right) + t_1 \Gamma\left(t_1 - t_1 \vartheta\right)\right] \left[g\left(\frac{nht_2 \vartheta}{\mu m d}\right)-1\right]\right\}. \label{eq:RedFact_dec_full}
\end{equation} 
Here, we introduced the dimensionless integration variable $\vartheta$ to simplify the according term in (\ref{eq:w_final}). We are are left with specifying the rate $\Gamma \left( t \right)$ and the decoherence function $g\left(s \right)$ for each independent decoherence process. In all but one case, the decoherence rate will be constant in time and (\ref{eq:RedFact_dec_full}) reduces to the simplified form 
Eq.~(6) given in the main text, with $f\left(s \right) = \int_0^1 \diff \vartheta g\left( s\vartheta \right) \in \left[0,1\right]$. The latter determines how strongly the decoherence process affects a finite interference path separation on average. The limit, where each scattering event takes away full which-way information,  is given by $f\left( s \right) = 0$, implying that the matter-wave coherence decays at the maximum rate $\Gamma$.

\subsubsection{Absorption of thermal radiation}

Each photon of frequency $\omega$ absorbed from the isotropic background radiation field transfers $\hbar\omega/c$ of momentum to the particle. The corresponding decoherence master equation for the reduced one/dimensional state of motion reads as \cite{hornberger2004theory}
\begin{equation}
 \cL_{\rm abs} \rho = \int \diff \omega \, \gamma_{\rm abs} \left( \omega \right) \left[ \int_{\left| \vn \right|=1} \frac{\diff^2 n }{4\pi} \exp\left( \frac{i\omega n_x \ox }{c} \right) \rho \exp\left( -\frac{i\omega n_x \ox }{c} \right) - \rho \right], \label{eq:ME_abs}
\end{equation} 
with $\gamma_{\rm abs} \left( \omega \right)$ the spectral absorption rate (\ref{eq:gabs}). This leads to the decoherence parameters
\begin{equation}
 \Gamma_{\rm abs} = \int\diff\omega \, \gamma_{\rm abs} \left( \omega \right), \quad g_{\rm abs}\left( x \right) = \int\frac{\gamma_{\rm abs} \left( \omega \right) \diff\omega}{\Gamma_{\rm abs}}\sinc \left( \frac{\omega x}{c} \right), \quad f_{\rm abs}\left( x \right) = \int\frac{\gamma_{\rm abs} \left( \omega \right) \diff\omega}{\Gamma_{\rm abs}} \, \frac{\Si\left( \omega x/c \right)}{\omega x/c}. \label{eq:dec_abs} 
\end{equation} 

\subsubsection{Emission of thermal radiation}

The emission of thermal radiation, the time-reversal of absorption, has the same effect on the motional state of the particle. Assuming an isotropic emission pattern, the same master equation (\ref{eq:ME_abs}) can be used to describe its impact, once the absorption rate is replaced by the spectral emission rate (\ref{eq:gemi}) for internally hot particles. If the internal temperature of the particle remains approximately constant the decoherence effect is described by parameters of the same form as (\ref{eq:dec_abs}). If however the particle does cool down significantly during flight, the emission rate will depend on time and the fringe reduction factor becomes
\begin{equation}
 R_n^{\rm (emi)} = \exp\left\{ \int\diff\omega\int_{0}^{1}\diff\vartheta\,\left[ t_{1}\gamma_{{\rm emi}}\left(\omega,T_{{\rm int}}\left(t_{1}-t_{1}\vartheta\right)\right)+t_{2}\gamma_{{\rm emi}}\left(\omega,T_{{\rm int}}\left(t_{1}+t_{2}\vartheta\right)\right)\right] \left[\sinc\left(\frac{nh\omega t_2 }{\mu m c d} \vartheta \right)-1\right]\right\}. \label{eq:dec_emi}
\end{equation} 
The time dependence of the internal temperature can be estimated by solving the differential equation $m c_m \partial_t T_{\rm int} = \int\diff\omega \, \hbar\omega \left[ \gamma_{\rm abs} \left( \omega \right) - \gamma_{\rm emi} \left( \omega,T_{\rm int} \right)\right]$, with $c_m$ the specific heat of the particle material. The solution is plotted in \ref{fig:heating} for the examples of silicon and silica nanoparticles.

\subsubsection{Elastic scattering of thermal radiation}

Rayleigh scattering of thermal background radiation is governed by the master equation
\begin{equation}
 \cL_{\rm sca}\rho = \int\diff\omega \, \gamma_{\rm sca} \left( \omega \right) \left\{ \int\frac{\diff^2 n \diff^2 n'}{16\pi^2} \exp\left[ \frac{i\omega \left( n_x - n_x' \right) \ox }{c} \right] \rho \exp\left[ \frac{i\omega \left( n_x' - n_x \right) \ox }{c} \right] - \rho  \right\}, \label{eq:ME_sca}
\end{equation} 
using the spectral scattering rate (\ref{eq:gsca}) and assuming once again an isotropic scattering pattern. This yields the decoherence parameters
\begin{equation}
  \Gamma_{\rm sca} = \int\diff\omega \, \gamma_{\rm sca} \left( \omega \right), \quad g_{\rm sca}\left( x \right) = \int\frac{\gamma_{\rm sca} \left( \omega \right) \diff\omega}{\Gamma_{\rm sca}} \sinc^2 \left( \frac{\omega x}{c} \right), \quad f_{\rm sca}\left( x \right) = \int\frac{\gamma_{\rm sca} \left( \omega \right) \diff\omega}{\Gamma_{\rm sca}} \left[ \frac{\Si\left( 2\omega x/c \right)}{\omega x/c}-\sinc^2\left( \frac{\omega x}{c} \right) \right]. \label{eq:dec_sca}
\end{equation}
Our simulations indicate that decoherence by Rayleigh scattering is negligible for the nanospheres considered here, but it will eventually be relevant for larger particles.

\subsubsection{Collisions with residual gas particles}

Collisions with residual gas particles transfers momentum in the same way as elastic light scattering. The corresponding master equation is therefore similar to (\ref{eq:ME_sca}). However, the decoherence function $g_{{\rm col}}\left( s \right)$ involves the particle-gas scattering amplitudes \cite{hornberger2004theory}, which we will not evaluate
here. Instead, we take the conservative estimate that each collision event transfers sufficiently many grating momenta on average that it fully resolves adjacent interference paths.
That is to say, the decoherence function vanishes for $s\neq0$ and the reduction factor simplifies to $R_n^{{\rm (col)}} = \exp\left[-\Gamma_{{\rm col}}\left( t_1 + t_2 \right)\right]$.

In order to estimate the effect, we assume that the residual gas of pressure $p_{g}$ consists of nitrogen ($m_g = 28\,$amu), in thermal equilibrium with the environment at temperature $T_{{\rm env}}$.
The mean velocity of the gas particles then reads as $v_{g}=\sqrt{2k_{B}T_{{\rm env}}/m_{g}}$. An approximate expression for the total scattering rate is obtained from a van der Waals scattering model \cite{hornberger2004theory},
\begin{equation}
\Gamma_{{\rm col}}\approx\frac{4\pi\Gamma\left(9/10\right)}{5\sin\left(\pi/5\right)}\left(\frac{3\pi C_{6}}{2\hbar}\right)^{2/5}\frac{p_{g}v_{g}^{3/5}}{k_{B}T_{{\rm env}}}, \qquad C_{6} \approx \frac{3\alpha\left( \omega=0 \right) \alpha_{g} I_{g}\, I}{32\pi^2\eps_0^2\left(I+I_{g}\right)}
\end{equation}
where $\Gamma$ denotes the Gamma function. The van der Waals coupling constant $C_6$ is estimated by means of the
London formula \cite{london1930theorie}, with $\alpha,\alpha_{g}$ the static polarizabilities and
$I,I_{g}$ the ionization energies of the nanospheres and the gas particles, respectively. For nitrogen, we use $\alpha_g = 1.74\,$\AA$^3 \times 4\pi\eps_0$ and $I_g = 15.6\,$eV \cite{crchandbook2010}. The ionization energy of the nanospheres is roughly estimated by the bulk work function, $I_{\rm Si},I_{\rm SiO2} \approx 5\,$eV; their static polarizabilities are computed using (\ref{eq:susceptibility}) and assuming static permittivities of $\eps_{\rm Si} = 11.9$ \cite{young1973compilation} and $\eps_{\rm SiO2} = 3.8$ (low-frequency value derived from the spectrum in \ref{fig:Spectra}).

\subsubsection{Overall fringe reduction}

Finally, the overall fringe reduction factor can be written as
\begin{eqnarray}
\ln \left( R_n \right) &=& -\Gamma_{{\rm col}}\left(t_{1}+t_{2}\right) + \int\diff\omega\,\gamma_{{\rm abs}}\left(\omega\right)\left[\frac{\Si\left(a_{n}\right)}{a_{n}}-1\right]\left(t_{1}+t_{2}\right) + \int\diff\omega\,\gamma_{{\rm sca}}\left(\omega\right)\left[\frac{\Si\left(2a_{n}\right)}{a_{n}}-\sinc^{2}\left(a_{n}\right)-1\right]\left(t_{1}+t_{2}\right) \nonumber \\
&& +\int\diff \omega \int_{0}^{1}\diff\vartheta\,\left\{ t_{1}\gamma_{{\rm emi}}\left[\omega,T_{{\rm int}}\left(t_{1}-t_{1}\vartheta\right)\right]+t_{2}\gamma_{{\rm emi}}\left[\omega,T_{{\rm int}}\left(t_{1}+t_{2}\vartheta\right)\right]\right\} \left[\sinc\left(a_{n}\vartheta\right)-1\right],\nonumber\\
\end{eqnarray}
where we introduced the abbreviation $a_{n}=n h \omega t_{2}/\mu m c d$ and Si is the sine integral.


The sinusoidal fringe visibility, 
Eq.~(5) in the main text, is then reduced by the factor $R_1$. It is plotted in \ref{fig:decoherence}(a) and (b) as a function of the total time of flight $t_1+t_2$ and the initial internal temperature $T_{\rm int} \left( 0 \right)$ for the $10^6\,$amu silicon and silica nanospheres, respectively. We assume a fixed background pressure of $p_g=10^{-10}\,$mbar, where collisional decoherence influences the visibility significantly only after approximately 500\,ms.

\section{Supplementary References}
\bibliography{bibliography}

\end{document}